\documentstyle[12pt,epsf,epsfig]{article}

\begin{document}

\begin{center}
{\bf A Quasicrystallic Domain Wall in Nonlinear Dissipative Patterns}

Boris A. Malomed\footnote{%
e-mail: malomed@eng.tau.ac.il}

Department of Interdisciplinary Studies, Faculty of Engineering, Tel Aviv
University, Tel Aviv 69978, Israel

\bigskip

Horacio G. Rotstein

Department of Chemistry and Volen Center for Complex Systems, Brandeis
University,

MS 015 Waltham, MA 02454-9110, USA
\end{center}

\newpage

\begin{center}
{\bf ABSTRACT}
\end{center}

We propose an indirect approach to the generation of a two-dimensional
quasiperiodic (QP) pattern in convection and similar nonlinear dissipative
systems where a direct generation of stable uniform QP planforms is not
possible. An {\it eightfold} QP pattern can be created as a broad transient
layer between two domains filled by square cells (SC) oriented under the
angle of $45$ degrees relative to each other. A simplest particular type of
the transient layer is considered in detail. The structure of the pattern is
described in terms of a system of coupled real Ginzburg-Landau (GL)
equations, which are solved by means of combined numerical and analytical
methods. It is found that the transient ``quasicrystallic'' pattern exists
exactly in a parametric region in which the uniform SC pattern is stable. In
fact, the transient layer consists of two different sublayers, with a narrow
additional one between them. The width of one sublayer (which locally looks
like the eightfold QP pattern) is large, while the other sublayer (that
seems like a pattern having a quasiperiodicity only in one spatial
direction) has a width $\sim 1$. Similarly, a broad stripe of a {\it %
twelvefold} QP pattern can be generated as a transient region between two
domains of hexagonal cells oriented at the angle of $30$ degrees.

\newpage

\section{Introduction}

Stable quasiperiodic (QP) planforms have been first discovered as
equilibrium patterns in metallic alloys \cite{quasi}. The simplest and most
important types of the QP planforms are the ten-, eight-, and twelvefold
ones (note an argument according to which only these types of the
two-dimensional quasicrystals may occur in physical systems \cite{Levitov}).
A $2n$-fold pattern is a superposition of $n\geq 4$ spatial harmonics based
on a set of equal-length wave vectors ${\bf k}_{j}$ with the equal angles $%
\pi /n$ between ${\bf k}_{j}$ and ${\bf k}_{j+1}$ (more general aperiodic
patterns with unequal wave vector lengths and/or unequal angles between the
vectors may also exist, but they have probably never been considered, except
for the case of unequilateral hexagons with $n=3$ \cite{Nuz}).

Then, it was predicted that the $2n$-fold structures of different particular
types may exist as dynamical planforms of a stationary form in
nonequilibrium systems, such as thermal convection and the like \cite{MNT}.
Later, the predicted pattern (a twelvefold one, in particular) was indeed
experimentally observed, first, as a stable dynamical nonequilibrium
planform in the Faraday ripples \cite{Jerry} (a large-aspect-ratio liquid
layer subject to high-frequency shaking in the vertical direction), and
recently in an optical cell filled with {\rm Na} vapors \cite{Logvin}.
Nevertheless, experimental generation of dynamical QP patterns is far from
being straightforward; in particular, in the work \cite{Jerry} it was
necessary to shake the liquid layer with a two-frequency quasiperiodic force
at a specially selected ratio between the two frequencies. These
experimental difficulties are related to the general fact that, in terms of
the corresponding coupled Ginzburg-Landau (GL) equations for amplitudes of
the spatial modes, a superposition of which gives rise to the pattern, a
formal solution for the QP planforms always exists, but it may be stable
only in a relatively narrow parametric region \cite{MNT}.

An objective of this work is to put forward a much easier possibility of
generating eight- and twelvefold QP planforms of a finite but large size (a
stripe) in ``normal'' systems, where direct generation of the
quasicrystallic patterns does not seem feasible. To this end, we notice that
the set of the wave vectors on which the eight- or twelvefold QP is based
may be regarded, in an obvious way, as a superposition of two half-sets of $2
$ or $3$ vectors, see Fig. \ref{vectors} below. Each half-set, in turn, may
give rise to a usual periodic pattern consisting of square cells (SC) or
hexagonal cells, respectively. Thus, a natural way to generate a stripe
filled with the QP pattern in a generic system (e.g., a convective layer),
in which a stable QP pattern is not available, but stable periodic SC and/or
hexagonal patterns do exist, is to produce it as a transient layer (``domain
wall'' \cite{MNT}) between two large domains filled with the periodic cells,
the (half-) sets of the two or three wave vectors in the two domains being
oriented under the angle, respectively, $45$ or $30$ degrees relative to
each other. The pattern in the transient layer will then be a superposition
of the two periodic patterns, having a full set of the four or six wave
vectors, respectively. The so generated stripe will feature a QP pattern,
provided that its width is essentially larger than the wavelength (a size of
the elementary cell in the corresponding periodic pattern). It is known that
the latter condition can be achieved, under special but not unrealistic
conditions, for domain walls in the nonequilibrium systems \cite{MNT}.

In fact, it may be simpler to generate, following this way, a twelvefold QP
pattern between two hexagonal domains, as it is usually much easier to find
a stable hexagonal structure than a stable SC one \cite{Busse}. However, the
theoretical analysis of the transient layer between two SC domains is much
simpler, therefore in this work we concentrate on the latter case. In any
case, it should be stressed that, although stable square-cell patterns are
rarer than their counterparts in the form of hexagons or rolls, examples of
stable square cellular planforms are known in gas flames \cite{square1},
thermal convection \cite{square2}, and in some optical systems \cite{square3}%
. The interest to the SC planforms has been recently revived by the
discovery of this pattern in a double-layer Marangoni convection (see the
e-print \cite{new} and references therein).

The rest of the paper is organized as follows. In section 2, we formulate
the model. Some analytical results, which predict, in particular, that the
transient layer between the two SC patterns has a complex structure,
consisting of {\em three} layers, two broad ones and a narrow sublayer
sandwiched between them, are obtained in section 3. In section 4, we display
results of a direct numerical solution of the stationary real GL equations,
which comply with the analytical predictions. The paper is concluded by
section 5.

\section{The Model}

\begin{figure}[ph]
\epsfig{file=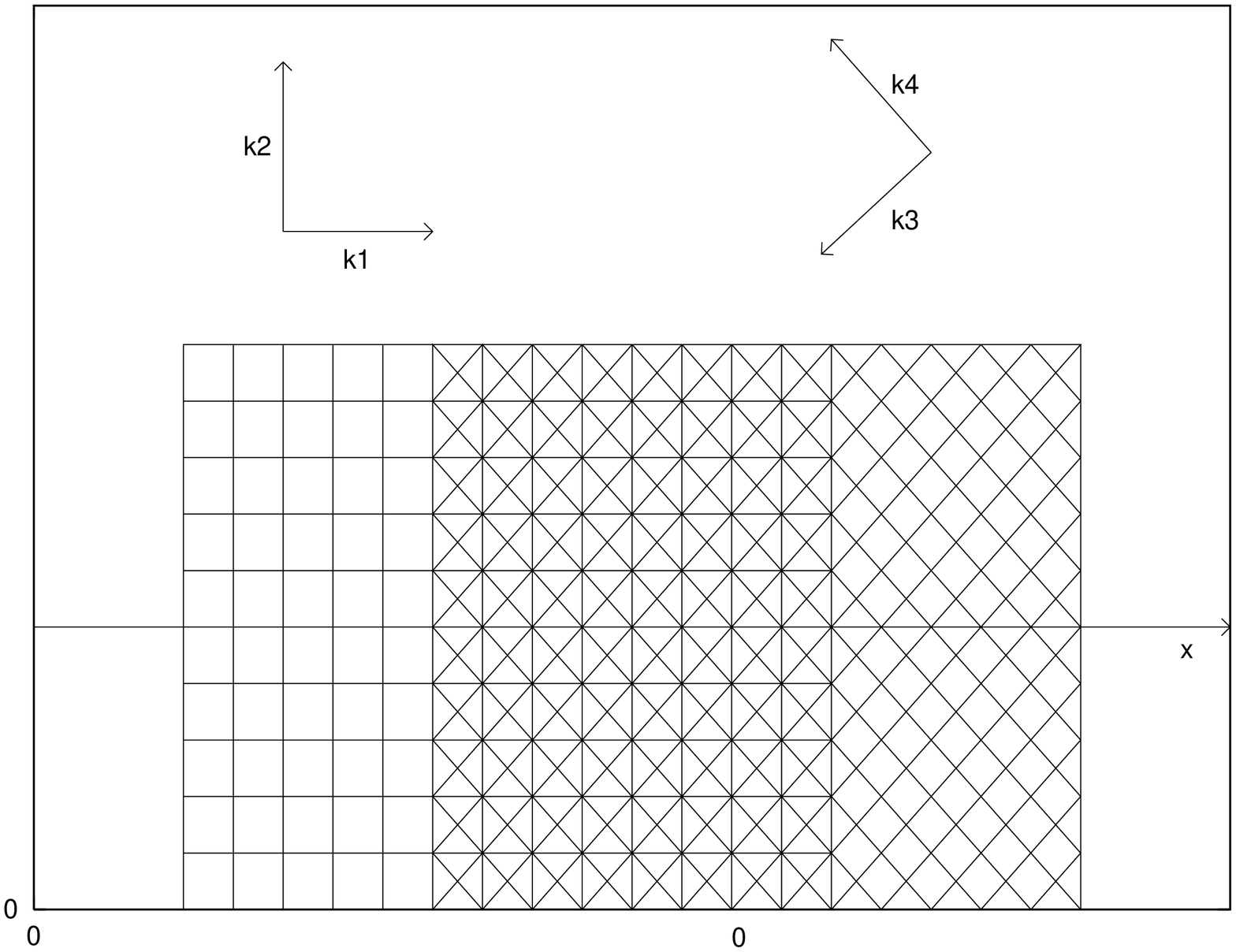,height=18cm,width=12cm,angle=-90}
\caption{}
\label{vectors}
\end{figure}

In this work, we assume a simplest configuration, shown in Fig. \ref{vectors}%
, that is going to give rise to the QP transient layer. The layer is
parallel to the $y$ axis, and the (half) sets of the two wave vectors are
chosen so that in the left domain the vectors ${\bf k}_{1,2}$ are parallel
to the $x$ and $y$ axes, while in the right domain both vectors ${\bf k}%
_{3,4}$ have the angle $45$ degrees relative to the axes. Accordingly, the
complex wave field describing the spatial distribution of physical variables
is assumed to be 
\begin{equation}
u({\bf r},t)=\sum_{j=1}^{4}B_{j}({\bf r},t)\,\exp \left( i{\bf k}_{j}{\bf r}%
\right) ,  \label{expansion}
\end{equation}
where{\bf \ }${\bf r}$ is the two-dimensional coordinate, and $B_{j}({\bf r}%
,t)$ are slowly varying amplitude functions.

In an experiments, the chosen configuration can be created, at least in a
part of the system, by means of specially selected boundary conditions,
which, in turn, are imposed by the sidewalls of the experimental cell.
Although we do not analyze the sidewall boundary conditions in this work, it
is obvious that, to support the configuration that we consider, one will
need to have a large-aspect-ratio cell with two sidewalls forming an angle $%
45$ degrees, which is quite possible. More general configurations, with
different orientations of the wave vectors relative to the layer between the
two domains, can be considered similarly to what is done below, but their
technical treatment will be more cumbersome.

A usual approach to the description of spatially nonuniform patterns of the
domain-wall type is based on a system of coupled real Ginzburg-Landau (GL)
equations for the slowly varying amplitudes, assuming that they do not
depend on the coordinate $y$ running along the stripe (wall), which is a
reasonable approximation if the stripe is long enough \cite{Pomeau,MNT2}.
Then, as it was shown in the mentioned works, the effective diffusion
coefficient (the one in front of the term $(B_{j})_{xx}$) in each GL
equation is 
\begin{equation}
D_{j}=\left( k_{j}^{(x)}\right) ^{2},  \label{D}
\end{equation}
where $k_{j}^{(x)}$ is the $x$-component of the vector. For the
configuration shown in Fig. \ref{vectors}, the system of the real GL
equations can be easily cast into a final form 
\begin{eqnarray}
A_{t} &=&(1/2)A_{xx}+A-A^{3}-2\left(
g_{2}A^{2}+g_{1}B_{1}^{2}+g_{1}B_{2}^{2}\right) A,  \label{A} \\
\left( B_{1}\right) _{t} &=&\left( B_{1}\right)
_{xx}+B_{1}-B_{1}^{3}-2\left( 2g_{1}A^{2}+g_{2}B_{2}^{2}\right) B_{1},
\label{B1} \\
\left( B_{2}\right) _{t} &=&B_{2}-B_{2}^{3}-2\left(
2g_{1}A^{2}+g_{2}B_{1}^{2}\right) B_{2},  \label{B2}
\end{eqnarray}
where $B_{1,2}$ are the amplitudes corresponding to the wave vectors{\bf \ }$%
{\bf k}_{1,2}$ in the left domain (Fig. \ref{vectors}), and, using the
obvious symmetry of the configuration shown in Fig. \ref{vectors}, we have
set $B_{3}=B_{4}\equiv A$.

Note that the diffusion coefficient in Eq. (\ref{B2}) is zero due to Eq. (%
\ref{D}). Generally speaking, in this case one should take into account a
higher-order derivative term $\sim \left( B_{2}\right) _{xxxx}$ \cite
{Pomeau,MNT2}. However, this is not necessary while the diffusion terms do
not vanish in the two other equations (see details below).

In Eqs. (\ref{A}) through (\ref{B2}), the linear gain coefficient and the
coefficient of the nonlinear self-interaction of the spatial mode are
normalized to be $1$, $g_{1}$ and $g_{2}$ being the coefficients of the
nonlinear interaction between the modes with the angles $\alpha =45$ and $90$
degrees between their carrier wave vectors. Normally, the nonlinear
interaction coefficient decreases with the increase of $\alpha $, so that 
\begin{equation}
g_{2}<g_{1}<1.  \label{inequalities}
\end{equation}
The necessary and sufficient stability conditions for the SC pattern are
well known \cite{MNT}, 
\begin{equation}
g_{2}<\frac{1}{2}\,,\,\,g_{1}>\frac{1}{4}\left( 1+2g_{2}\right) \,,
\label{stability}
\end{equation}
while the conditions providing for stability of the eightfold QP planform
are 
\begin{equation}
g_{2}<\frac{1}{2}\,,\,\,g_{1}<\frac{1}{4}\left( 1+2g_{2}\right) \,.
\label{QPstability}
\end{equation}
An obvious feature of the two sets of the stability conditions is their
incompatibility, i.e., SC and QP can never be stable simultaneously.

In fact, the inequality $g_{2}<1/2\,$\ in the set of the conditions (\ref
{stability}) is a cause for the relative rarity of stable SC planforms, as,
despite the general property (\ref{inequalities}), the actual dependence $%
g(\alpha )$ is usually weak, so that $g_{2}$ is not essentially smaller than 
$1$. Nevertheless, the full set of the SC stability conditions (\ref
{stability}) can be satisfied in the above-mentioned physical systems.

A consequence of the conditions (\ref{QPstability}) necessary for the
stability of the QP pattern is $g_{1}<1/2$, which, with regard to Eq. (\ref
{inequalities}), makes the stability of the QP pattern still less feasible
than that of the SC one. An objective of this work is to propose a way to
produce a transient QP pattern between two {\em stable} SC domains with
different orientations (Fig. \ref{vectors}) in the case when \ the uniform
QP planform is unstable. Note that a similar approach is known as a way to
generate of a broad stripe of a SC pattern between two domains of
orthogonally oriented rolls in the case when the uniform SC pattern is
unstable, while the rolls are stable \cite{MNT2}.

In the case when the SC stability conditions (\ref{stability}) are met, it
is quite reasonable to assume that the coefficients $g_{1,2}$ are close to $%
1/2$ (because it is physically implausible to have $g_{1,2}$ much smaller
than $1/2$), which suggests to present them as 
\begin{equation}
g_{1,2}\equiv \frac{1}{2}\left( 1-\mu _{1,2}\right) ,\,|\mu _{1,2}|\ll 1.
\label{mu}
\end{equation}
As it will be seen below, the smallness of $\mu _{1,2}$ naturally provides
for the strip of the QP pattern to be broad, which is exactly the condition
justifying the consideration of the transient-layer patterns. Note that $\mu
_{2}$ must be positive according to Eq. (\ref{stability}), while $\mu _{1}$
may formally have either sign. However, it will be shown below that the
necessary solution does not exist if $\mu _{1}<0$.

Below, we will also use a parameter 
\begin{equation}
m\equiv 2\mu _{1}/\mu _{2},  \label{m}
\end{equation}
which is, generally, $\sim 1$. In terms of $m$, the second stability
condition (\ref{stability}) takes a very simple form, $m<1$.

Eqs. (\ref{A}), (\ref{B1}), and (\ref{B2}) must be supplemented by boundary
conditions (b.c.) to guarantee that, at $x\rightarrow \pm \infty $ (recall
the system is formally assumed to be infinitely large), the pattern
considered asymptotically coincides with either of the two SC planforms
composed of the modes $B_{1}$ and $B_{2}$ or $B_{3}=B_{4}\equiv A$.
Obviously, this implies 
\begin{eqnarray}
\lim_{x\rightarrow +\infty }A(x) &=&1/\sqrt{2-\mu _{2}}\equiv A_{\lim
},\,\lim_{x\rightarrow +\infty }B_{1,2}(x)=0,  \label{+infinity} \\
\lim_{x\rightarrow -\infty }A(x) &=&0,\,\lim_{x\rightarrow -\infty
}B_{1,2}(x)=A_{\lim }\,.  \label{-infinity}
\end{eqnarray}
Formally, this set may seem overdetermined, as we add six b.c. to the
fourth-order system of Eqs. (\ref{A}) and (\ref{B1}) (Eq. (\ref{B2})
contains no $x$-derivatives). However, it is easy to check that the
seemingly superfluous b.c. for $B_{2}$ are nothing else but direct
corollaries of the four legitimate b.c. for $A$ and $B_{1}$.

\section{Analytical Results}

A stationary version ($\partial /\partial t=0$) of Eqs. (\ref{A}), (\ref{B1}%
) and (\ref{B2}) can be essentially simplified, as in this case Eq. (\ref{B2}%
) becomes just an algebraic relation, that has two solutions: $B_{2}=0$, or 
\begin{equation}
B_{2}^{2}=1-\left( 2-m\mu _{2}\right) A^{2}-\left( 1-\mu _{2}\right)
B_{1}^{2},  \label{eliminate}
\end{equation}
$m$ being the parameter defined by Eq. (\ref{m}). First, we consider the
case when the expression (\ref{eliminate}) holds (obviously, it may hold as
long as it yields $B_{2}^{2}>0$). Substituting it into the stationary
versions of Eqs. (\ref{A}) and (\ref{B1}), we obtain 
\begin{eqnarray}
\mu _{2}^{-1}A^{\prime \prime }+mA-\left[ \left( 2\left( 2m-1\right)
-m^{2}\mu _{2}\right) A^{2}+\left( 2-m\mu _{2}\right) B_{1}^{2}\right] A
&=&0,  \label{Astat} \\
\mu _{2}^{-1}B_{1}^{\prime \prime }+B_{1}-\left[ \left( 2-\mu _{2}\right)
B_{1}^{2}+\left( 2-m\mu _{2}\right) A^{2}\right] B_{1} &=&0,  \label{B1stat}
\end{eqnarray}
the prime standing for $d/dx$. If, instead of Eq. (\ref{eliminate}), we take 
$B_{2}=0$, the stationary equations take the form 
\begin{eqnarray}
A^{\prime \prime }+2A-\left[ 2\left( 2-\mu _{2}\right) A^{2}+\left( 2-m\mu
_{2}\right) B_{1}^{2}\right] A &=&0,  \label{Astat0} \\
B_{1}^{\prime \prime }+B_{1}-\left[ B_{1}^{2}+\left( 2-m\mu _{2}\right) A^{2}%
\right] B_{1} &=&0.  \label{B1stat0}
\end{eqnarray}
It is noteworthy that, although the SC patterns may be stable at $\mu _{1}<0$%
, i.e., $m<0$, Eq. (\ref{Astat}) cannot have a solution for $A(x)$
exponentially decaying at $x\rightarrow -\infty $ (see the b.c. (\ref
{-infinity})) unless $m>0$, hence the present problem has {\em no solution}
with $m<0$.

Obviously, a solution to Eqs. (\ref{Astat}) and (\ref{B1stat}) can satisfy,
with regard to Eq. (\ref{eliminate}), the b.c. (\ref{-infinity}). However,
the same set of equations (\ref{Astat}) and (\ref{B1stat}) {\em cannot}
satisfy the b.c. (\ref{+infinity}): setting $B_{1}=0$ and $A={\rm const}$,
one obtains from Eq. (\ref{Astat}) 
\begin{equation}
A^{2}=A_{0}^{2}\equiv \frac{m}{2\left( 2m-1\right) -m^{2}\mu _{2}}\,\,,
\label{A0}
\end{equation}
which is obviously different from the necessary limit value $A_{\lim
}^{2}\equiv 1/\left( 2-\mu _{2}\right) $. Moreover, the value of $B_{2}^{2}$
corresponding to $A_{0}^{2}$ as per Eq. (\ref{eliminate}) is different from
zero, which also violates the b.c. (\ref{+infinity}).

In fact, the stationary state corresponding to the asymptotic value (\ref{A0}%
) with $B_{2}\neq 0$ is another uniform pattern, which is a superposition of
three spatial harmonics. This pattern is periodic in one spatial direction
and quasiperiodic in the other one. As it was demonstrated in \cite{MNT},
this pattern is always dynamically unstable, hence a solution having it as
an asymptotic state is physically irrelevant.

However, a solution to Eqs. (\ref{Astat}) and (\ref{B1stat}) makes sense as
long as it provides for $B_{2}^{2}>0$ according to Eq. (\ref{eliminate}).
Taking the asymptotic state $A^{2}=A_{0}^{2}$, $B_{1}^{2}=0$, one finds that
it gives rise to {\em negative} $B_{2}^{2}$ exactly in the case $m<1$, which
is considered in this work, as this is the case when the SC pattern is
stable, see above. Thus, Eqs. (\ref{Astat}) and (\ref{B1stat}) should be
used in the region $-\infty <x<x_{0}$, where, by definition, $x=x_{0}$ is a point at
which $B_{2}^{2}\left( x\right) $, as given by Eq. (\ref{eliminate}),
vanishes, i.e., 
\begin{equation}
\left( 2-m\mu _{2}\right) A^{2}(x_{0})+\left( 1-\mu _{2}\right)
B_{1}^{2}(x_{0})=1.  \label{x0}
\end{equation}
At the point $x=x_{0}$, one must switch from Eqs. (\ref{Astat}) and (\ref{B1stat}%
) to Eqs. (\ref{Astat0}) and (\ref{B1stat0}), setting $B_{2}\equiv 0$ at $%
x\geq x_{0}$. The continuity dictates to take the values 
$A(x\rightarrow x_{0}^{-})$
and $B_{1}(x\rightarrow x_{0}^{-})$ as the b.c. to Eqs. (\ref{Astat0}) and (\ref
{B1stat0}) at $x=x_{0}$, the b.c. at $x=+\infty $ being fixed by Eqs. (\ref
{+infinity}).

The transition from Eqs. (\ref{Astat}) and (\ref{B1stat}) to Eqs. (\ref
{Astat0}) and (\ref{B1stat0}) at $x=x_{0}$ provides for the continuity of all
the functions $A(x)$ and $B_{1,2}(x)$, but their first derivatives suffer a
jump at $x=x_{0}$. In fact, if the above-mentioned fourth-derivative term is
added to Eq. (\ref{B2}), the jump of the derivative will be smoothed down in
a narrow boundary layer. It should be stressed that the presence of the jump
does not violate the applicability of the description in terms of the GL
equations; the only problem is the absence of a detailed description of the
narrow boundary layer in which the jump is smoothed by the fourth-order
derivative term, but details of the inner structure of the narrow layer do
not affect the global picture.

Thus, the transient layer between the two SC domains consists of two
sublayers, described, respectively, by Eqs. (\ref{Astat}) and (\ref{B1stat}%
), and by Eqs. (\ref{Astat0}) and (\ref{B1stat0}), with the discontinuity of
the derivative between them. The first (left) sublayer contains all the four
spatial harmonics, hence is locally seems as a QP pattern. An important
finding is that, in the case of small $\mu _{2}$ considered here, the width
of this sublayer is large, scaling $\sim \mu _{2}^{-1/2}$, according to Eqs.
(\ref{Astat}) and (\ref{B1stat}). The possibility to produce a broad QP
stripe justifies all the consideration of the transient layer. The second
(right) sublayer contains only three spatial harmonics, as $B_{2}\equiv 0$
in it, hence it locally looks like a pattern quasiperiodic only in one
direction \cite{MNT}), and its width is $\sim 1$ (i.e., not specifically
large), according to Eqs. (\ref{Astat0}) and (\ref{B1stat0}).

This qualitative analysis of the transient layer's structure will be
corroborated and illustrated by direct numerical results displayed in the
next section. However, before using the numerical methods, one can notice
that the structure of the second (right) sublayer can be described in an
approximate analytical form if we consider a special case when the value $%
A(x=x_{0})$ at the internal boundary between the two sublayers is already close
to the asymptotic value $A_{\lim }$, and, accordingly, the value $B_{1}^{2}$
is small (in other words, this is case when $0<1-m\ll 1)$. Then, we may set 
\begin{equation}
A(x)\equiv A_{\lim }-a(x),  \label{a}
\end{equation}
where $a(x)$ is positive but small and vanishes at $x=+\infty $. Using the
smallness of $a(x)$ and $B_{1}$, one can simplify Eqs. (\ref{Astat0}) and (%
\ref{B1stat0}): 
\begin{eqnarray}
(1/\sqrt{2})a^{\prime \prime }-\left[ \sqrt{2}(2-\mu _{2})a-B_{1}^{2}\right]
-\mu _{1}B_{1}^{2} &=&0,  \label{simplea} \\
B_{1}^{\prime \prime }+\left[ \sqrt{2}(2-\mu _{2})a-B_{1}^{2}\right] B_{1}-%
\frac{1}{2}\left( \mu _{1}-2\mu _{2}\right) B_{1} &=&0.  \label{simpleB1}
\end{eqnarray}
The b.c. for $a(x)$ at $x=x_{0}$ can be obtained from the expansion of the exact
b.c. (\ref{x0}): 
\begin{equation}
a(x_{0})=\frac{B_{1}^{2}(x_{0})}{\sqrt{2}\left( 2-\mu _{2}\right) }\,.
\label{simplex0}
\end{equation}

The lowest-order approximate analytical solution to Eqs. (\ref{simplea}) and
(\ref{simpleB1}), satisfying the necessary boundary conditions, is very
simple: 
\begin{equation}
B_{1}(x)=B_{1}(x_{0})\,\exp (-\lambda x),\,\,a(x)=a(x_{0})\,\exp (-2\lambda x),
\label{a_solution}
\end{equation}
with $\lambda =\sqrt{\left( \mu _{2}-2\mu _{1}\right) /2}$. Note that the
condition necessary for $\lambda $ to be real is again exactly tantamount to 
$m<1$.

\section{Numerical Results}

To check the qualitative predictions for the structure of the transient
layer obtained in the previous section, we performed a two-stage numerical
integration of Eqs. (\ref{Astat}) and (\ref{B1stat}), and then of Eqs. (\ref
{Astat0}) and (\ref{B1stat0}). To this end, the first pair of the equations
was solved with the b.c. (\ref{-infinity}), continuing the solution until it
hits a point where it satisfies Eq. (\ref{x0}) (recall the latter condition
is equivalent to the vanishing of $B_{2}^{2}$). Actually, the numerical
integration was performed, instead of the original coordinate $x$, in terms
of a variable $\xi \equiv \tanh x$. This transformation is convenient
because it maps the semi-infinite intervals $(\pm \infty ,0)$ of the
variable $x$ into finite ones $(\pm 1,0)$.

The values $A(x_{0})$ and $B_{1}(x_{0})$, obtained from the solution of Eqs. (\ref
{Astat}) and (\ref{B1stat}), where then used to find a solution to Eqs. (\ref
{Astat0}) and (\ref{B1stat0}) in the interval $\tanh x_{0}<\xi <1$, satisfying the
b.c. (\ref{+infinity}) at $\xi =1$. In accord with what said above, a
condition of the continuity of the first derivatives across the point $\xi
=\tanh x_{0} $ was not imposed.

\begin{figure}[ph]
\begin{tabular}{llllllll}
{\large {\bf (a) }} & \epsfig{file=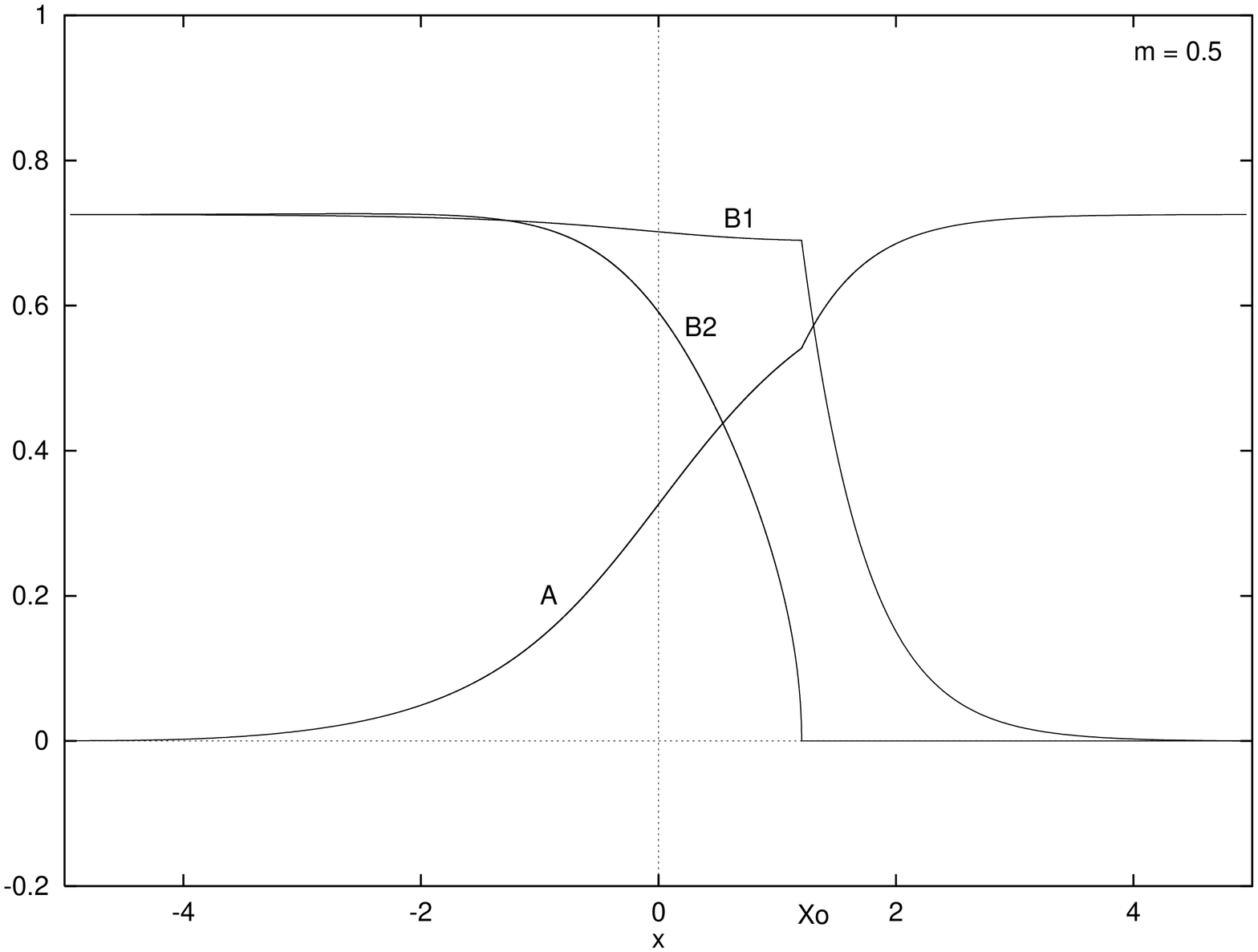,height=9cm,width=6cm,angle=-90}
&  &  &  &  &  &  \\ 
{\large {\bf (b) }} & \epsfig{file=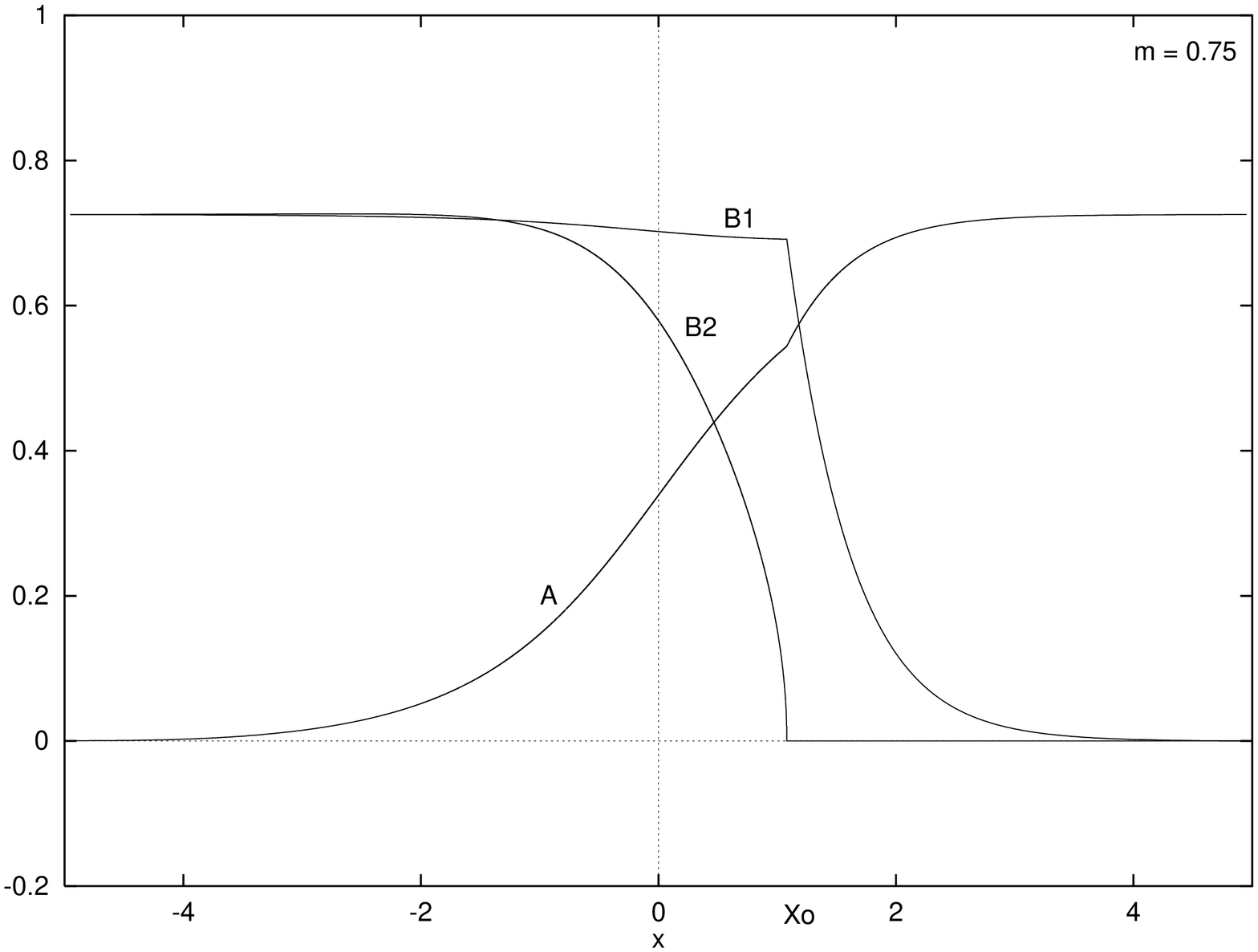,height=9cm,width=6cm,angle=-90}
&  &  &  &  &  &  \\ 
{\large {\bf (c) }} & \epsfig{file=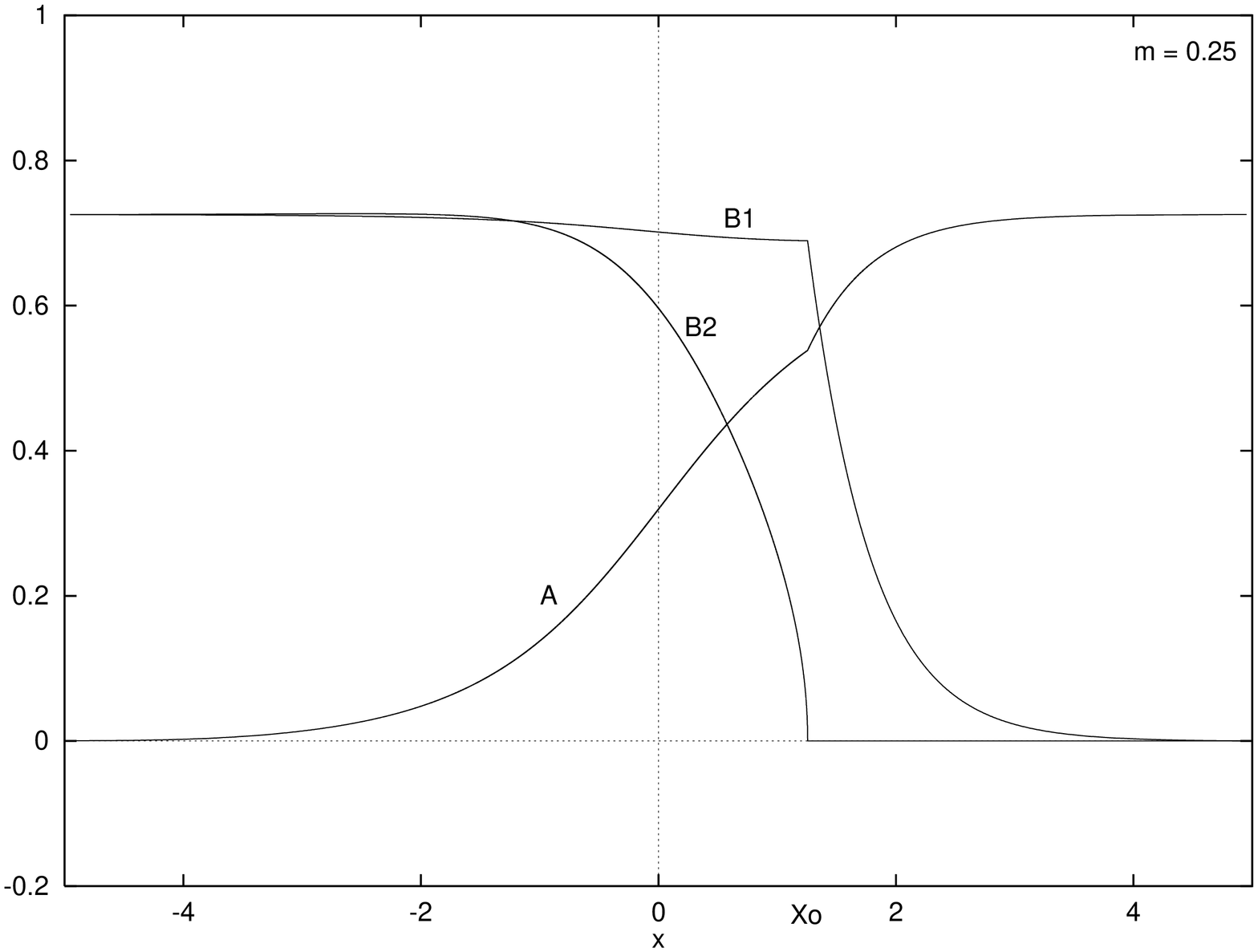,height=9cm,width=6cm,angle=-90}
&  &  &  &  &  & 
\end{tabular}
\caption{}
\label{numerica}
\end{figure}

Three typical examples of the thus obtained numerical solutions are
displayed vs. the coordinate $x$ in Fig. \ref{numerica}, for three
characteristic values $m=0.75$, $m=0.5$, and $m=0.25$, the parameter $\mu
_{2}$, that must be small enough, being fixed in all the three cases as $\mu
_{2}=0.1$. A characteristic feature clearly seen in all the cases is that,
in accord with the prediction of the above analysis, a width of the left
sublayer is essentially larger than that of the right one. It is also
noteworthy that the solutions changes very little with a large change in $m$%
, i.e., the transient layer between the two SC domains is expected to be
quite robust.

\section{Conclusion}

In this work, we have proposed an approach that makes it possible to the
generate a two-dimensional quasiperiodic pattern in nonlinear dissipative
systems where a direct generation of stable uniform quasiperiodic planforms
is not possible. An eightfold pattern can be created in the form of a broad
transient stripe between two domains filled by square cells, which are
oriented under the angle of $45$ degrees relative to each other. Using the
symmetry of the configuration considered, the structure of the pattern was
described in terms of a system of three coupled real stationary
Ginzburg-Landau equations, which were analyzed by means of analytical and
numerical methods. It was found that the transient quasiperiodic pattern
exists exactly in a parametric region in which the uniform square-cell
pattern is stable. Further, it was found that the transient layer consists
of two different sublayers, with a derivative jump between them (that can be
smoothed into an additional narrow boundary layer, if higher-order
derivatives are added to the Ginzburg-Landau equations). The width of the
sublayer that features the eightfold quasiperiodic pattern is found to be
large, while the other sublayer (filled with a less interesting pattern,
which is quasiperiodic only in one direction) has a width $\sim 1$. A broad
stripe of a {\it twelvefold} QP pattern can be similarly generated as a
transient layer between two domains of hexagonal cells oriented at the angle
of $30$ degrees.

It still remains to perform simulations of the full time-dependent
Ginzburg-Landau equations, in order to directly test the dynamical stability
of the broad transient layers. However, the numerically found robustness of
the layers against the variation of the crucial control parameter $m$
suggests that they have a good chance to be dynamically stable.

\newpage

\section*{Figure Captions}

Fig. 1. The configuration giving rise to the transient layer filled with the
eightfold quasi-periodic pattern between two domains of square cells
oriented under the angle $45$ degrees relative to each other..

Fig. 2. Numerically found structure of the transient layer corresponding to
the configuration shown in Fig. 1. The small parameter $\mu _{2}$ defined by
Eq. (\ref{mu}) is fixed to be $0.1$, while the control parameter $m$,
defined by Eq. (\ref{m}), takes values $0.75$ (a), $0.50$ (b), and $0.25$
(c) (recall only the values $0<m<1$ make sense in the present context).

\newpage

\end{document}